\documentclass[12pt]{article}
\usepackage{amsmath}
\usepackage{amssymb}
\usepackage{bm}
\newcommand{\be}{\begin{equation}}
\newcommand{\ee}{\end{equation}}

\newcommand{\bea}{\begin{eqnarray}}
\newcommand{\eea}{\end{eqnarray}}

\newcommand{\mbold}[1]{\mbox{\boldmath$#1$}}

\begin{document}
\title{%
\large\bf NUCLEAR INCOMPRESSIBILITY IN THE\\
 QUASILOCAL DENSITY FUNCTIONAL THEORY}
\author{%
V. B. Soubbotin and V. I. Tselyaev\\
{\it \normalsize
 Nuclear Physics Department,
 V. A. Fock Institute of Physics,}\\
{\it \normalsize
 St. Petersburg State University, 198504,
 St. Petersburg, Russia}\\$\vphantom{,}$\\
 X. Vi\~nas\\
{\it \normalsize
 Departament d'Estructura i Constituents de la Mat\`eria,}\\
{\it \normalsize
 Facultat de F\'{\i}sica, Universitat de Barcelona,}\\
{\it \normalsize
 Diagonal 645 E-08028 Barcelona, Spain}}

\date{March 19, 2004}
\maketitle

\begin{abstract}

We explore the ability of the recently established quasilocal density
functional theory for describing the isoscalar giant monopole resonance.
Within this theory we use
the scaling approach and perform constrained calculations for obtaining
the cubic and inverse energy weighted moments (sum rules) of the RPA strength.
The meaning of the sum rule approach in this case is discussed.
Numerical calculations are carried out using Gogny forces and an excellent
agreement is found with HF + RPA results previously reported in literature.
The nuclear matter compression modulus predicted in our model
lies in the range 210-230 MeV which agrees with earlier findings.
The information provided by the sum rule approach in the case of nuclei near
the neutron drip line is also discussed.

\vspace{2em}
\begin{flushleft}
PACS numbers: 21.60Jz, 31.15Ew, 31.15Gy
\end{flushleft}

\end{abstract}
\newpage

\section{\large INTRODUCTION}

Recently we have established the quasilocal density functional theory
(QLDFT) and its application for describing the nuclear
ground state properties \cite{STV}. It is just a generalization of the
local Hohenberg-Kohn-Sham (HKS) theory \cite{HK,KS} and is based
on the definition of an universal energy density functional ${\cal E}$ of
the Slater density matrix (DM) $\rho_0$
\cite{STV}. One can define the quasilocal energy functional
$E[\hat{n},\hat{\tau},\hat{\mbold J}]$ through a many-to-one mapping
of the density matrix $\rho_0$ to a set of local quantities
$\hat{n}\equiv\{n_p,n_n\}$, $\hat{\tau}\equiv\{\tau_p,\tau_n\}$,
$\hat{\mbold J}\equiv\{{\mbold J}_p,{\mbold J}_n\}$,
where $n_q$, $\tau_q$, ${\mbold J}_q$
are the local, the kinetic energy, and the spin densities of
each kind of nucleons ($q=p,n$):
\be E[\hat{n},\hat{\tau},\hat{\mbold J}]
=\inf_{\rho_0\to \hat{n},\hat{\tau},\hat{\bf J}}
{\cal E}[\rho_0]\,.
\label{eq1} \ee
The main property of the functional (\ref{eq1}) is
that its minimum provides the exact ground state energy, which is attained at
the true local nucleon densities $\hat{n}$. It is worthwhile to note that
in this case the equilibrium
$\hat{\tau}$ and $\hat{\mbold J}$ densities are not the exact
ones and they
correspond to the system without correlations.

The functional $E[\hat{n},\hat{\tau},\hat{\mbold J}]$ consists of
two terms:
\be
E[\hat{n},\hat{\tau},\hat{\mbold
J}]=E_0[\hat{n},\hat{\tau},\hat{\mbold J}]+E_{RC}[\hat{n}],
\label{eq2} \ee
where $E_0[\hat{n},\hat{\tau},\hat{\mbold J}]$ is a Hartree-Fock
(HF) energy functional reduced to the quasilocal form,
and $E_{RC}$ is the residual correlation energy.
The HF contribution corresponds to a finite-range
density-independent effective force.
Its quasilocal form is calculated using the
extended Thomas-Fermi theory \cite{SV}, and consists of a kinetic energy part,
a Hartree term, a local exchange
contribution, and a spin-orbit energy as the
used in the Skyrme energy functionals \cite{STV}.
To compute this HF functional we use the
density-independent
finite-range part of the Gogny force \cite{Gog}.
The main formulas for $E_0$ are reported
in \cite{STV}. The residual correlation energy $E_{RC}[\hat{n}]$
is taken phenomenologically and is parametrized
similar to the contribution of
the density-dependent part of the Skyrme or the Gogny interactions.

The aim of the QLDFT is to describe
properties of the nuclear ground state such as the total
energy $E_{GS}$, the local particle densities $\hat{n}$, the neutron and proton
separation energies (chemical potentials) of double magic nuclei. To
do that use is made of the variational principle to obtain a set of
single-particle equations similar to the Kohn-Sham ones but containing
additionally a position-dependent effective mass and a spin-orbit potential
\cite{STV}. We have used this quasilocal approach to obtain some nuclear
ground-state properties of several magic nuclei with the Gogny D1S force
 \cite{D1S}. We have found a very good agreement between the binding
energies and root mean square radii computed using the QLDFT and the
corresponding values calculated with the full HF method \cite{STV}.

Now we are in turn to investigate if the QLDFT is also able to predict
some information about excited nuclear states.
In this paper we want to analyze the
collective breathing mode and to study the isoscalar giant monopole
resonance (ISGMR) in finite nuclei using our model.
The experimental value of the excitation energy of the ISGMR is
mainly extracted from the
analysis of the inelastic $(\alpha, \alpha')$
scattering data (see \cite{YCL} and references therein). From the
experimental excitation energies
in medium and heavy nuclei it is possible to estimate the nuclear matter
compression modulus $K_{\infty}$. In particular, self-consistent HF plus
random phase approximation (RPA)
calculations using Gogny \cite{Bl} or Skyrme forces \cite{Giai} determine
$K_{\infty}$ to be 210-230 MeV.

\section{\large THE BASIC THEORY}

Giant resonances are understood in terms of small amplitude oscillations of
nuclei as a response to an external field generated by electromagnetic
or hadronic probes.
The most widely used
theoretical framework for describing these
vibrations is the RPA \cite{RS} which allows to obtain the strength function
$S(E)$ that measures the nuclear response.
For medium and heavy nuclei far from the drip line, the strength $S(E)$
corresponding to the breathing mode is mainly concentrated in a rather narrow
region of the energy spectrum.
In these cases the knowledge of a few low energy
moments of $S(E)$ (sum rules), which are defined as
\be
m_k = \sum_{\nu \neq 0} (E_{\nu} - E_0)^k
| \langle \nu | Q | 0 \rangle |^2 \,,
\label{defmk} \ee
where $Q=\sum_i r_i^2$ is the one-body monopole
excitation operator, can provide a
useful information on the average properties of the ISGMR. If $k$ is an odd
integer, the sum rules $m_k$ can be written as expectation values of some
commutators calculated in the exact ground state \cite{Boh}. For example
\be
m_1 = < 0 \vert [Q,[H,Q]] \vert 0>
\label{eq4} \ee
and
\be
m_3 = < 0 \vert [[Q,H],[H,[H,Q]]] \vert 0>.
\label{eq5} \ee
From these moments one can estimate several average energies as:
\be
\bar{E}_{k,k-2}=\sqrt{\frac{m_k}{m_{k-2}}}\,.
\label{eq6} \ee

The full quantal calculation of the sum rules is still a complicated task
because the exact ground state wave function is usually unknown.
However, if the energy
moments of $S(E)$ are evaluated at 1p1h (RPA) level, it is possible to replace
the exact ground-state wave functions by the uncorrelated HF ones
in the calculation of the sum rules.

For forces which commutes with the monopole
excitation operator $Q$, as it happens
for the Skyrme and Gogny interactions, the only contribution to the commutator
$[Q,[H,Q]]$ comes from the kinetic energy part of the Hamiltonian. Thus the
$m_1$ sum rule is given by (see \cite{Boh})
\be
m_1=\frac{2\hbar^2}{M}A<r^2>_0\,,
\label{defm1}\ee
where $M$ is the nucleon mass,
the expectation value of the operator $r^2$ is calculated with the HF
wave functions.

The direct evaluation of the commutators
entering the formula (\ref{eq5}) for $m_3$ is a rather cumbersome
task which can be avoided computing the $m_3$ sum rule of the RPA monopole
strength function through the scaling method \cite{Boh}. Using the scaled
ground-state wave function
\be
\Psi_{\eta}({\mbold r}_1, \ldots , {\mbold r}_A) =
\eta^{\frac{3}{2}A} \Psi(\eta {\mbold r}_1, \ldots , \eta {\mbold r}_A)\,,
\label{eq8}\ee
the plus three energy moment can be expressed by means of the second
derivative of the scaled ground-state energy
$E^s_{GS}(\eta)=<\Psi_{\eta}|H|\Psi_{\eta}>$ as
\be
m_3=\frac{1}{2}\left(\frac{2\hbar^2}{M}\right)^2
\left(\frac{d^2 E^s_{GS}(\eta)}
{d \eta^2}\right)_{\eta=1}.
\label{defm3} \ee
 Thus the $m_3$ moment measures the change of energy of the nucleus when the
ground-state wave function is
deformed according to (\ref{eq8}).

 With account of Eqs.~(\ref{eq6}), (\ref{defm1}), and
(\ref{defm3}),
the scaled average energy of the ISGMR is defined as:
\be
\bar{E}_{GMR}^s = \sqrt{\frac{m_3}{m_1}}\,.
\label{eq10} \ee
This method allows also
to define the scaled nuclear compression modulus for a finite
nucleus with mass number $A$ as (see \cite{Blaiz})
\be
K_A^s=\frac{M}{\hbar^2} <r^2>_0 (\bar{E}_{GMR}^s)^2 =
 \frac{1}{A}\left(\frac{d^2 E^s_{GS}(\eta)}
 {d \eta^2}\right)_{\eta=1}.
\label{defka}\ee

Let us describe the method of calculation of the derivatives in the
right-hand sides of Eqs. (\ref{defm3}) and (\ref{defka}).
Within the QLDFT the scaled energy can be written as
\be
E^s_{GS}(\eta)=T(\eta)+E_{Nucl}^{Dir}(\eta)+E_{Nucl}^{Exch}(\eta)
+E_{Coul}(\eta)+E_{so}(\eta)+E_{RC}(\eta).
\label{eqEeta} \ee
Due to the fact that in the QLDFT we deal with Slater determinant wave functions,
the different contributions to the scaled
energy can be easily determined. The kinetic, the spin-orbit, and the Coulomb
contributions scale as:
\be
T(\eta)=\eta^2T(1), \qquad
E_{so}(\eta)=\eta^5E_{so}(1), \qquad {\rm and} \qquad  E_{Coul}(\eta)=\eta
E_{Coul}(1).
\label{TSOC} \ee
In our approach the residual correlation energy
is chosen phenomenologically (see Introduction) as:
\be
E_{RC}[\hat{n}]= \frac{t_3}{4} \int d{\mbold r} n^{\alpha}({\mbold
r}) [(2+x_3) n^2({\mbold r}) - (2 x_3 + 1) \left( n_p^2({\mbold
r}) + n_n^2({\mbold r}) \right)],
\label{ERXC}\ee
which under the scaling transformation reads:
\be
E_{RC}(\eta)=\eta^{3(\alpha+1)}E_{RC}(1).
\ee

These formulas give a simple way to obtain some of the terms
entering (\ref{eqEeta}). Similar explicit expressions
for the direct and exchange energy contributions coming from the finite range
part of the effective force cannot be derived. However, one can calculate
their corresponding derivatives with respect to the
dimensionless collective coordinate
$\eta$ using the following trick. For a sake of simplicity we will
consider a simple Wigner central force. In our approach we use
Gogny-like forces with Gaussian form factors $v(r/\mu)$
where $\mu$ is the range of the force. In this case the direct
part of the energy coming from the finite range effective force
has the general form
\be
E_{Nucl}^{Dir}=\frac{1}{2} \int d{\mbold r}d{\mbold r}'n({\mbold
r})v(|{\mbold r}-{\mbold r}'|/\mu)n({\mbold r}')\,.
\ee
It is easy to see that the scaling of the direct energy is reduced
to renormalization of the range $\mu$:
\be
E_{Nucl}^{Dir}(\eta)= \frac{1}{2} \int
d{\mbold r}d{\mbold r}'n({\mbold r})v(|{\mbold r}-{\mbold
r}'|/(\eta\mu))n({\mbold r}')=\tilde{E}_{Nucl}^{Dir}(\mu)\,.
\label{edir}
\ee
It enables us to write (see \cite{KA,PVCE})
\be
\left.
\frac{d^2 E_{Nucl}^{Dir}(\eta)}{d\eta^2}\right|_{\eta=1} =
\frac{1}{2} \int d{\mbold r}d{\mbold r}' n({\mbold r})
\left[2 s \frac{d v(s/\mu)}{d s} +
s^2 \frac{d^2 v(s/\mu)}{d s^2} \right] n({\mbold r}'),
\ee
where $s= {\mbold r} - {\mbold r}'$ is the relative coordinate.
In our calculations we, however, used another relation:
\be
\left.
\frac{d^2 E_{Nucl}^{Dir}(\eta)}{d\eta^2}\right|_{\eta=1}=\mu^2 \left.
\frac{d^2\tilde{E}_{Nucl}^{Dir}(\mu')}{d\mu'^2}\right|_{\mu'=\mu},
\label{d2ed}
\ee
which follows immediately from Eq.~(\ref{edir}).

The same is true for the exchange Fock energy $E_{Nucl}^{Exch}(\eta)$
coming from the finite range effective force. Our quasilocal
exchange energy is obtained by replacing the exact Slater
single-particle density matrix
by the quasi-classical DM within the extended Thomas-Fermi approximation
(ETF) \cite{SV}. It is easy to check that the
quasiclassical DM $\rho_{ETF}({\mbold r},{\mbold r}')
= \rho_{ETF}({\mbold R},{\mbold s})$ (with ${\mbold R}= ({\mbold
r}+{\mbold r}')/2$) satisfies
the correct scaling transformation law:
\be
\rho_{ETF}({\mbold R},{\mbold s}, \eta)=\eta^3\rho_{ETF}({\eta\mbold
R},\eta{\mbold s})\,
.\ee
Thus, because in the QLDFT the exchange energy has the general form \be
E_{Nucl}^{Exch}= \frac{1}{2} \int d{\mbold R}d{\mbold s}\rho^2_{ETF}({\mbold
R},{\mbold s})v(s/\mu)\,,\ee
we have
\be E_{Nucl}^{Exch}(\eta)=\int d{\mbold R}d{\mbold s}
\rho^2_{ETF}({\mbold R},{\mbold s})v(s/(\eta\mu))=
\tilde{E}_{Nucl}^{Exch}(\mu)\,,
\ee
and
\be
\left.
\frac{d^2E_{Nucl}^{Exch}(\eta)}{d\eta^2}\right|_{\eta=1}=\mu^2
\left.
\frac{d^2\tilde{E}_{Nucl}^{Exch}(\mu')}{d\mu'^2}\right|_{\mu'=\mu}\,.
\label{d2ee}
\ee
The derivatives in Eqs. (\ref{d2ed}), (\ref{d2ee}) were calculated numerically.

In light nuclei the strength function is much more spread and fragmented
than in heavy nuclei (see for instance \cite{YLC}).
Therefore, to get more insight about the spreading of
$S(E)$ we will consider another estimate of the ISGMR mean energy provided by
a constrained QLDFT calculation.
It is well known (see \cite{Boh}) that
the $m_{-1}$ RPA sum rule is half the ground-state polarizability with
respect to the excitation operator $Q$ of the ISGMR, i.~e.
\be
m_{-1}= \sum_{\nu \neq 0}
\frac{| \langle \nu | Q | 0 \rangle |^2}{E_{\nu} - E_0}
= -\frac{1}{2}\left(\frac{d R^2(\lambda)}{d\lambda}\right)_{\lambda=0}
= \frac{1}{2}\left(\frac{d^2 E^c_{GS}(\lambda)}{d\lambda^2}\right)_{\lambda=0},
\label{defmm1}\ee
where $R^2(\lambda)$ and $E^c_{GS}(\lambda)$ are the expectation values of the
excitation ($Q$) and the Hamiltonian ($H$) operators
evaluated with the ground-state wave function
of the constrained Hamiltonian $H_c = H + \lambda Q$. Using Eqs. (\ref{eq6}),
(\ref{defm1}), and (\ref{defmm1}),
the constrained estimate of the average energy of the ISGMR
is defined as
\be
\bar{E}_{GMR}^c = \sqrt{\frac{m_1}{m_{-1}}}\,.
\ee
The constrained nuclear compression modulus can also be defined in a similar
way to that of Eq.~(\ref{defka}):
\be
K_A^c=\frac{M}{\hbar^2} <r^2>_0 (\bar{E}_{GMR}^c)^2 =
\frac{1}{A} \left(R^2\frac{d^2 \tilde{E}^c_{GS}(R)}{d
R^2}\right) _{R=R_0},
\label{defkac}\ee
 where the function $\tilde{E}^c_{GS}(R)$ is defined by the relation
$\tilde{E}^c_{GS}(R(\lambda))=E^c_{GS}(\lambda)$, $R_0=R(\lambda=0)$.

As it has been pointed out before, in Ref. \cite{Boh} was proved that the
exact ground-state wave function can be replaced by the HF one in the
self-consistent HF + RPA (1p1h) sum rule calculation. Analogous considerations
allow to state that the $m_3$, $m_1$, and $m_{-1}$ sum rules calculated
within QLDFT (i.e. using Slater determinants
built up of the single-particle wave functions
obtained from the minimization of the QLDFT energy functional),
coincide with the QLDFT-based self-consistent RPA results.

Finally, let us note that
the RPA moments fulfill $\sqrt{m_3/m_1} \ge m_1/m_0 \ge \sqrt{m_1/m_{-1}}\,.$
Therefore, the scaled ($\bar{E}_{GMR}^s$)
and the constrained ($\bar{E}_{GMR}^c$) estimates of the average energy of the
resonance give an upper and a lower bound of the mean energy of the ISGMR
$m_1/m_0$.
The total RPA width $\sigma$
of the strength distribution can also be estimated
from the $\bar{E}_{GMR}^s$ and $\bar{E}_{GMR}^c$ energies:
\be
\sigma \le \frac{1}{2} \sqrt{(\bar{E}_{GMR}^s)^2 - (\bar{E}_{GMR}^c)^2}.
\label{width} \ee

\section{\large NUMERICAL RESULTS}

We have calculated the scaled and the constrained estimates of the average
energies of the ISGMR as well as the corresponding compression moduli
in finite nuclei
using the QLDFT with the D1$'$ \cite{Gog} and the D1S \cite{D1S} Gogny forces.
In order to test our method we compare in Table~\ref{tab1} our
QLDFT results
with the calculations performed in Ref.~\cite{Bl} in the framework of
the HF and HF + RPA approaches with the same forces.
\begin{table}[h]
\begin{center}
\caption{\label{tab1}
The ISGMR average energies (in MeV) in $^{208}$Pb computed in the
framework of the scaled ($\bar{E}_{GMR}^s$) and the constrained
($\bar{E}_{GMR}^c$) approaches with the D1$'$ and the D1S Gogny
forces using the QLDFT compared with the average energies obtained
in Ref.~\cite{Bl} using the HF and the HF + RPA methods.}
\vspace{3mm}
\tabcolsep=2.1em
\renewcommand{\arraystretch}{1.5}%
\begin{tabular}{lccccc}
\hline
\hline
\multicolumn{1}{c}{}& \multicolumn{2}{c}{QLDFT}&
\multicolumn{1}{c}{HF}&
\multicolumn{2}{c}{HF + RPA}\\
 & $\bar{E}_{GMR}^s$ & $\bar{E}_{GMR}^c$ & $\bar{E}_{GMR}^c$
 & $\bar{E}_{1,\,-1}$ & $\bar{E}_{3,\,1}$ \\
\hline
 D1$'$ & 14.51 & 13.93 & 14.05 & 14.15 & 15.33 \\
 D1S   & 13.65 & 13.05 & 13.22 & 13.34 & 14.16 \\
\hline
\hline
\end{tabular}
\end{center}
\end{table}
It can be seen that the agreement between the constrained QLDFT and
HF energies $\bar{E}_{GMR}^c$, and the average HF + RPA energies
$\bar{E}_{1,\,-1}$ is fairly well. The agreement between the scaled
QLDFT energies $\bar{E}_{GMR}^s$ and the values of $\bar{E}_{3,\,1}$
obtained in HF + RPA approach is reasonable,
though it is worse than for the constrained energies.
From these comparisons we conclude that using
our sum rule approach based on the QLDFT, the corresponding average
energies can be confidently used to theoretically estimate the ISGMR
energies with Gogny forces, at least for nuclei for which the
monopole strength shows a well defined narrow peak.

In Table~\ref{tab2} we display results obtained for the nuclei
$^{40}$Ca, $^{90}$Zr, and $^{208}$Pb, for which the experimental
excitation energies of the ISGMR are accurately known \cite{YCL,YLC}.
\begin{table}[h]
\begin{center}
\caption{\label{tab2}
The compression moduli $K_A$ (in MeV) and the ISGMR average
energies $\bar{E}_{GMR}$ (in MeV) of some magic nuclei computed in
the framework of the scaled ($K_{A}^s$, $\bar{E}_{GMR}^s$) and the
constrained ($K_{A}^c$, $\bar{E}_{GMR}^c$) approaches with the
D1$'$ and the D1S Gogny forces using the QLDFT compared with the
same quantities computed with the SkM$^{*}$ and the SIII Skyrme
interactions using the HF approximation. Experimental average
energies are taken from Refs. \cite{YCL,YLC}.} \vspace{3mm}
\tabcolsep=0.85em
\renewcommand{\arraystretch}{1.5}%
\begin{tabular}{llccccc}
\hline
\hline
 && $^{16}$O & $^{28}$O & $^{40}$Ca & $^{90}$Zr & $^{208}$Pb \\
\hline
  $K_{A}^s$ &
    D1$'$   & 132 & 106 & 146 & 154 & 152 \\
  & D1S     & 126 & 100 & 137 & 141 & 137 \\
  & SkM$^*$ & 124 & 103 & 138 & 146 & 143 \\
  & SIII    & 199 & 166 & 225 & 243 & 243 \\
  $\bar{E}_{GMR}^s$ &
    D1$'$   & 28.1 & 20.1 & 23.2 & 19.0 & 14.5 \\
  & D1S     & 27.1 & 19.3 & 22.2 & 18.0 & 13.7 \\
  & SkM$^*$ & 26.8 & 19.6 & 22.2 & 18.3 & 13.9 \\
  & SIII    & 34.5 & 25.3 & 28.5 & 23.4 & 17.9 \\
\hline
  $K_{A}^c$ &
    D1$'$   & 102 & 14 & 127 & 144 & 140 \\
  & D1S     & 100 & 14 & 121 & 132 & 125 \\
  & SkM$^*$ &  94 & 28 & 119 & 136 & 133 \\
  & SIII    & 132 & 17 & 178 & 216 & 214 \\
  $\bar{E}_{GMR}^c$ &
    D1$'$   & 24.7 &  7.2 & 21.6 & 18.4 & 13.9 \\
  & D1S     & 24.1 &  7.3 & 20.9 & 17.5 & 13.1 \\
  & SkM$^*$ & 23.2 & 10.2 & 20.6 & 17.6 & 13.3 \\
  & SIII    & 28.1 &  8.1 & 25.4 & 22.1 & 16.8 \\
\hline
  $\bar{E}_{GMR}^{exp}$
  & $\hphantom{19.2 \pm}$
  & $\hphantom{19.2 \pm 0.4}$
  & $\hphantom{19.2 \pm 0.4}$
  & 19.2 $\pm$ 0.4
  & 17.9 $\pm$ 0.2
  & 14.2 $\pm$ 0.3 \\
\hline
\hline
\end{tabular}
\end{center}
\end{table}
For the nucleus $^{208}$Pb, the QLDFT scaled and constrained
predictions of the ISGMR excitation energy computed with the D1$'$ Gogny force
agree very well with the experimental centroid value (14.2 MeV).
On the other hand the QLDFT results obtained with the D1S force underestimate
the experimental value
by a 3.5 \% ($\bar{E}_{GMR}^s$) and 8 \% ($\bar{E}_{GMR}^c$).
The reason for this discrepancy lies on the fact
that the nuclear matter compression modulus
for the D1$'$ force ($K_{\infty}=$ 228 MeV) is larger than that for
the D1S force ($K_{\infty}=$ 209 MeV).
The situation is inverse for the $^{90}$Zr nucleus where the
theoretical QLDFT estimates of the ISGMR energy
for D1S force are in a better agreement
with the experimental centroid value (17.9 MeV)
than that for the D1$'$ force.
Thus $K_{\infty}$ extracted from the sum rule approach using the
results for the nucleus $^{208}$Pb is around 230 MeV, while the
results for the nucleus $^{90}$Zr lead to the value around 210
MeV, which corresponds to the the upper and lower bounds of
the compression
modulus predicted in Ref. \cite{Bl}. For the $^{40}$Ca nucleus,
the ISGMR energies obtained with our QLDFT plus sum rule approach
clearly overestimate the experimental centroid value both for the
D1$'$ and for the D1S forces. In particular, for the D1S force the
calculated scaled energy overestimates the $(m_3/m_1)^{1/2}$
experimental value \cite {YLC} by a 7~\% whereas the constrained
energy is larger than the experimental $(m_1/m_{-1})^{1/2}$ by a
21~\%. For this nucleus our QLDFT model predicts a total RPA
width
(Eq.~(\ref{width})) of 3.7 MeV. This means that the monopole
strength is quite spread or fragmented and thus the sum rule
estimate of the ISGMR for $^{40}$Ca should be considered only as
qualitative.

The use of radioactive beams allows to obtain nuclei beyond the limits of
$\beta$-stability and to reach the neutron drip line for nuclei with $Z \le 8$.
Nuclei near the drip line are characterized by the small energy of the
last bound levels and by their large asymmetry. It is expected that the
properties of such nuclei, in particular the ones concerning collective
excitations, may considerably differ from the corresponding properties of
stable nuclei. Theoretical HF + RPA calculations carried out by
Hamamoto, Sagawa, and Zhang
\cite{HS1} using the SkM$^{*}$ force, have shown that in nuclei near
the drip line the distribution of the monopole strength is much affected by the
presence of the low-energy threshold due to the tiny bound nucleons. This
pattern clearly differs from the one shown by stable nuclei where the strength
is concentrated in a single peak at least for medium and heavy nuclei.

In particular the aforementioned authors have analyzed the monopole strength
of the $^{28}$O nucleus \cite{HS2}. We have also considered this nucleus in our
QLDFT plus sum rule approach and their scaled and constrained energies are
also reported in the Table~\ref{tab2}. Although the sum rule
approach is unable to deal with fine details of the monopole strength,
it provides some signatures about the global behaviour of the
monopole strength in nuclei near the drip line. In the nucleus $^{28}$O, the
scaled and the
constrained energies are clearly reduced with respect to the corresponding
values
in the stable $^{16}$O nucleus, which are also given in the Table~\ref{tab2}. This
reduction is particularly significant
for the constrained energy and produces a
noticeable enhancement of the width
of the ISGMR. These facts point out the
importance of the low-energy part of the monopole strength in drip line
nuclei.

In order to do a complementary check of our QLDFT estimates of the average
excitation energies of the ISGMR using the D1$'$ and the D1S forces,
we have repeated the sum
rule analysis using the Skyrme force SkM$^{*}$, which has $K_{\infty}$=217 MeV
slightly larger than the corresponding D1S value (209 MeV). The scaled
energies with Skyrme forces are easily calculated \cite{Boh} and the
constrained calculations are performed in the way discussed previously.
The excitation energies as well as the finite nuclei
incompressibilities obtained with SkM$^{*}$ are also collected in the Table~\ref{tab2}.
It can be seen that D1S and SkM$^{*}$ practically predict the same values for
the scaled and constrained energies. Actually the excitation energies obtained
with SkM$^{*}$ are slightly larger than the ones computed with D1S, at
least for medium and heavy nuclei where the sum rule approach can be regarded
more confidently, due to the fact that $K_{\infty}$ is a little
bit larger
in the former than in the latter force. In the same Table~\ref{tab2} we also report the
ISGMR excitation energies obtained using
the Skyrme SIII force which has $K_{\infty}$=355
MeV. As it has been pointed out in \cite{Bl}, the energy of the breathing mode
is roughly proportional to $\sqrt{K_{\infty}}$. We have checked this
behaviour with the excitation energies of the ISGMR of the nuclei
$^{90}$Zr and $^{208}$Pb calculated with the D1$'$, D1S, SkM$^*$, and SIII
forces using the scaling method and performing constrained calculations. It is
found that this proportionality is also reasonably well fulfilled for each
nucleus when the QLDFT values are used, especially in the case of the scaled
energies.

\section{\large SUMMARY}

We have analyzed the ability of the recently established quasilocal density
functional theory for describing the breathing mode in finite nuclei.
First of all, it has to be pointed out that
the 1p1h RPA calculations based on this theory have similar
properties to the HF + RPA calculations. In particular, the exact RPA
ground-state wave function can be replaced by the
Slater determinant built up of the single-particle QLDFT
wave functions in order to obtain the
odd moments of the 1p1h RPA strength function. Thus the scaling approach and
constrained calculations can be used to obtain the $m_3$ and $m_{-1}$ moments
(sum rules) of the RPA strength based on the quasilocal density functional
theory.

We have applied this sum rule approach using the Gogny forces. For the nucleus
$^{208}$Pb, we have found an excellent agreement with
the HF and the HF + RPA results for the ISGMR
reported in the literature. Using the D1$'$ and the D1S Gogny forces and
comparing our
theoretical estimates of the average energies of the ISGMR with the experimental
values of some selected medium and heavy nuclei, we have found that the
predicted
nuclear matter compression modulus lies in the range 210-230 MeV which
also agrees with earlier findings. It is also found that our estimates of the
average energies of the ISGMR, in particular the ones obtained with the
scaling method,
roughly scale with the square root of the
compression modulus in nuclear matter.

Although the sum rule approach works basically for medium and heavy stable
nuclei
where the RPA strength is mainly concentrated in a narrow peak, we have
analyzed its predictions
in nuclei near the neutron drip line. In this case the
RPA strength is broaded and the low energy part considerably enhanced because
of the last weakly bound neutrons. The sum rule approach is able to give
globally this tendency in such nuclei by reducing the $m_3$ and especially the
$m_{-1}$ sum rules and thereby increasing the estimate of the resonance width.

\vspace{3em}
The authors are grateful to Prof. S.~Shlomo for useful discussions
and Dr. M. Centelles for a careful reading of the manuscript.
V.~B.~S. and V.~I.~T. would like to acknowledge financial support from
the Russian Ministry of Education under grant E02-3.3-463.
X.~V. acknowledges financial support from DGI (Ministerio de Ciencia y
Tecnolog\'{\i}a and FEDER (Spain) under grant
BFM2002-01868 and from DGR (Catalonia) under grant 2001SGR-00064.
\newpage

\end{document}